\documentclass{anstrans}
\title{Production of Heavy Clusters with an Expanded Coalescence Model in CEM}
\author{Leslie M. Kerby,$^{1,2}$ Stepan G. Mashnik$^{1}$}

\institute{
$^{1}$Los Alamos National Laboratory, Los Alamos, New Mexico USA
\and
$^{2}$University of Idaho, Idaho Falls, Idaho USA
}

\email{leslie31415@gmail.com}

\usepackage{graphicx} % allows inclusion of graphics
\usepackage{booktabs} % nice rules (thick lines) for tables
\usepackage{microtype} % improves typography for PDF

\usepackage[nottoc,numbib]{tocbibind}	%LMK
\settocbibname{References}
\usepackage{gensymb}

\begin{document}
%%%%%%%%%%%%%%%%%%%%%%%%%%%%%%%%%%%%%%%%%%%%%%%%%%%%%%%%%%%%%%%%%%%%%%
\section{Introduction}
The production of heavy clusters in nuclear reactions is important in a wide variety of applications: radiation shielding, space engineering and design, medical physics, accelerator design, and more. According to the Cascade Exciton Model 
(CEM)~\cite{CEMModel, Trieste08}, there are three ways high-energy heavy clusters can be produced. The first way is via coalescence of nucleons produced in the IntraNuclear Cascade (INC). The second way is via the preequilibrium model. The last way is via Fermi breakup. Previous work in CEM examines the impact of expansions of the preequilibrium model and Fermi breakup model on heavy cluster production~[3--7]. The present work studies the impact of expanding the coalescence model on heavy cluster spectra. 
CEM03.03~\cite{Trieste08}, the default event generator in the Monte Carlo N-Particle transport code version 6 (MCNP6)~\cite{MCNP6} for intermediate-energy nuclear reactions, is capable of producing light fragments up to $^4$He in its coalescence model. In the present study, we have expanded the coalescence model to be able to produce up to $^7$Be. Preliminary results are promising. 

%%%%%%%%%%%%%%%%%%%%%%%%%%%%%%%%%%%%%%%%%%%%%%%%%%%%%%%%%%%%%%%%%%%%%%
\section{Background}
When the cascade stage of a reaction is completed, CEM uses the coalescence model described in Ref. \cite{Toneev1983, KKG+Schulz-coal1983} to ``create'' high-energy d, t, $^3$He, and $^4$He by final-state interactions among emitted cascade nucleons outside of the target nucleus. The coalescence model used in CEM is similar to other coalescence models for heavy-ion-induced reactions. The main difference is that instead of complex-particle spectra being estimated simply by convolving the measured or calculated inclusive spectra of nucleons with corresponding fitted coefficients, CEM03.03 uses in its simulations of particle coalescence real information about all emitted cascade nucleons and does not use integrated spectra. (Note that the coalescence introduced recently in the Li\`ege intranuclear cascade (INCL) [11-14], is in a way similar to the coalescence considered by CEM as proposed in Ref. \cite{Toneev1983, KKG+Schulz-coal1983}, with the main contrast being that INCL considers coalescence of INC nucleons on the border of a nucleus, inside the target-nucleus, while CEM coalesces INC nucleons outside the nucleus.) We assume that all the cascade nucleons having differences in their momenta smaller than $p_c$ and the correct isotopic content form an appropriate composite particle. The coalescence radii $p_c$ as used in CEM03.03 are:\\
Incident energy, $T$, $< 300$ MeV or $> 1000$ MeV
\begin{eqnarray}
p_c(d) & = & 90 \mbox{ MeV/c ;} \nonumber \\
p_c(t) & = & p_c(^3{\mbox He}) = 108 \mbox{ MeV/c ;} \\
p_c(^4{\mbox He}) & = & 115 \mbox{ MeV/c .} \nonumber 
\label{eq:coalescence}
\end{eqnarray}
$300$ MeV $< T < 1000$ MeV
\begin{eqnarray}
p_c(d) & = & 150 \mbox{ MeV/c ;} \nonumber \\
p_c(t) & = & p_c(^3{\mbox He}) = 175 \mbox{ MeV/c ;} \\
p_c(^4{\mbox He}) & = & 175 \mbox{ MeV/c .} \nonumber
\label{eq:coalescence2}
\end{eqnarray}

If several cascade nucleons are chosen to coalesce into composite particles, they are removed from the distributions of nucleons and do not contribute further to such nucleon characteristics as spectra, multiplicities, {\it etc}.

%%%%%%%%%%%%%%%%%%%%%%%%%%%%%%%%%%%%%%%%%%%%%%%%%%%%%%%%%%%%%%%%%%%%%%%%%%%%%%%%
\section{Coalescence Expansion}
The Coalescence Model in CEM03.03 allows for coalescence up to $^4$He. We have expanded this to additionally allow for the coalescence of $^6$He, $^6$Li, $^7$Li, and $^7$Be. 

CEM03.03 uses the simplest version of the coalescence model
\cite{Toneev1983, KKG+Schulz-coal1983} and checks only the momenta of nucleons
emitted during the cascade stage of reactions, without checking their coordinates.

The momentum, $p$, of each nucleon is calculated 
relativistically
from its kinetic energy, $T$ 
(CEM03.03 provides in its output files the energy of particles, but not their 
momenta).
%, as in Eq.~\ref{KE_Mom}: 
%\begin{equation}
%p^2 c^2 = KE(KE+2m_0 c^2),
%\label{KE_Mom}
%\end{equation}
%where $m_0$ is the rest mass of the nucleon. Eq.~\ref{KE_Mom} can be derived from the %%%relativisitc energy relations%
%\begin{equation}
%E^2 = p^2 c^2 + m_0^2 c^4
%\label{eq:E1}
%\end{equation}
%and
%\begin{equation}
%E = KE + m_0 c^2.
%\label{eq:E2}
%\end{equation}
Coalescence occurs if each nucleon in the group has $|\Delta p| \le p_c$, where $\Delta p$ is defined as the difference between the nucleon momentum and the average momentum of all nucleons in the group.

The coalescence model of CEM03.03 first checks all nucleons to form 2-nucleon pairs, their momenta permitting. It then checks if an alpha particle can be formed from two 2-nucleon pairs (either from 2 n-p pairs or from a n-n and p-p pair). After this it checks to see if any of the 2-nucleon pairs left can combine with another nucleon to form either tritium or $^3$He. And lastly, it checks to see if any of these 3-nucleon groups (tritium or $^3$He) can coalesce with another nucleon to form $^4$He. 

The expanded coalescence model then takes these 2-nucleon pairs, 3-nucleon (tritium or $^3$He only) groups, and $^4$He to see if they can coalesce to form heavy clusters. $^4$He can coalesce with a 3-nucleon group to form either $^7$Be or $^7$Li. Two 3-nucleon groups can coalesce to form either $^6$Li or $^6$He. And $^4$He can coalesce with a 2-nucleon pair to form either $^6$Li or $^6$He. All coalesced nucleons are removed from the distributions of nucleons so that our coalescence model conserves both atomic- and mass-numbers. In addition, this expansion requires an insignificant amount 
(2--3 \%)
of increased computation time.

For additional details of the Coalescence Model expansion, see Ref.~\cite{CoalescenceLANL}.

\subsection{Coalescence Parameter $p_c$}
As mentioned 
%in the background section, 
above,
$p_c$ determines how dissimilar the momenta of nucleons can be and still coalesce. $p_c$ was expanded to also include a value for heavy clusters, or light fragments (LF): $p_c(LF)$. Our new $p_c$'s for incident energies, $T$, less than 300 MeV or greater than 1000 MeV are:
\begin{eqnarray}
p_c(d) & = & 90 \mbox{ MeV/c ;} \nonumber \\
p_c(t) & = & p_c(^3{\mbox He}) = 108 \mbox{ MeV/c ;} \\
p_c(^4{\mbox He}) & = & 130 \mbox{ MeV/c .} \nonumber \\
p_c(LF) & = & 175 \mbox{ MeV/c .} \nonumber 
\label{eq:exp_coalescence1}
\end{eqnarray}

And for $300$ MeV $< T < 1000$ MeV
\begin{eqnarray}
p_c(d) & = & 150 \mbox{ MeV/c ;} \nonumber \\
p_c(t) & = & p_c(^3{\mbox He}) = 175 \mbox{ MeV/c ;} \\
p_c(^4{\mbox He}) & = & 205 \mbox{ MeV/c .} \nonumber \\
p_c(LF) & = & 250 \mbox{ MeV/c .} \nonumber 
\label{eq:exp_coalescence2}
\end{eqnarray}

%Please 
Note that the $p_c(^4{\mbox He})$ was also increased compared to the old $p_c$ values. Too many alpha particles were lost (coalesced into heavy clusters), and therefore we compensated by increasing the coalescence of $^4$He.  

%%%%%%%%%%%%%%%%%%%%%%%%%%%%%%%%%%%%%%%%%%%%%%%%%%%%%%%%%%%%%%%%%%%%%%%%%%%%%%%%
\section{Results and Analysis}
Examples of 
%preliminary 
results of our coalescence expansion are displayed in Figs.~1--3.
%\ref{fig:p1200Au} and~\ref{fig:p480Ag}. 
\begin{figure}[b]
\centering
\includegraphics[width=3.1in]{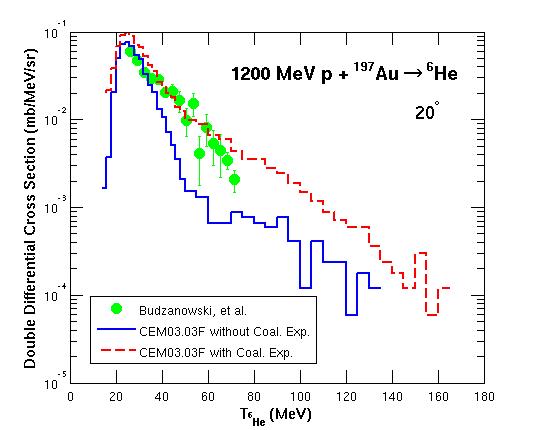}
\includegraphics[width=3.1in]{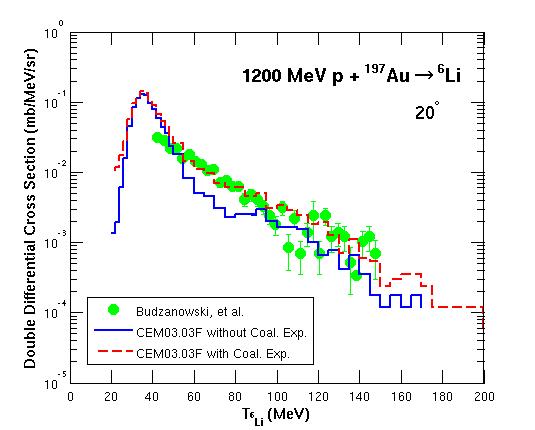}
\caption[]{Comparison of experimental data by Budzanowski, {\it et al.} \cite{Budzanowski} (green circles) for the production of $^6$He and $^6$Li at an angle of 20\degree ~from the reaction 1200 MeV p + $^{197}$Au, with results from CEM03.03F without coalescence expansion (blue solid lines) and CEM03.03F with coalescence expansion (red dashed lines).}
\label{fig:p1200Au}
\end{figure}
The upgraded CEM03.03F {\bf without} coalescence expansion (blue solid lines) and the upgraded CEM03.03F {\bf with} coalescence expansion (red dashed lines) are compared with experimental data (green circles). This coalescence expansion analysis is part of a larger project~[3--7]
%\cite{NIMB2015,FY2014,NIMA2014,ANS2014,ND2013} 
aimed at producing high-energy light fragments in spallation reactions. CEM03.03F refers to the upgraded CEM03.03 code, which has been upgraded with a Modified-Exciton-Model expansion and a total reaction cross section model improvement. The blue solid lines contain both of these improvements over the original CEM03.03, and the red dashed lines contain both of these improvements plus the coalescence expansion.

\begin{figure}[t]
\centering
\includegraphics[width=3.1in]{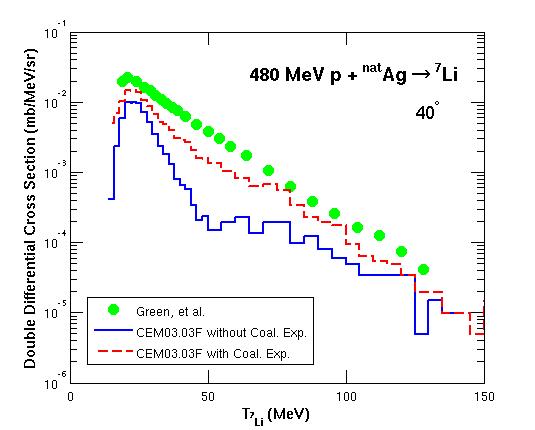}
\includegraphics[width=3.1in]{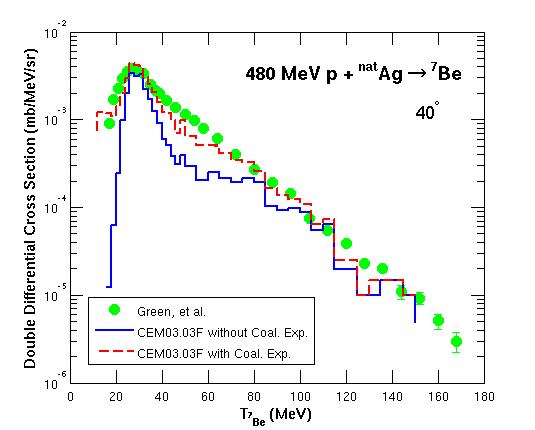}
\caption[]{Comparison of experimental data by Green {\it et al.} \cite{Green480} (green circles) for the production of $^7$Li and $^7$Be at an angle of 40\degree ~from the reaction 480 MeV p + $^{nat}$Ag, with results from CEM03.03F without coalescence expansion (blue solid lines) and CEM03.03F with coalescence expansion (red dashed lines).}
\label{fig:p480Ag}
\end{figure}

Fig.~\ref{fig:p1200Au} displays fragment production spectra of $^6$He and $^6$Li for the reaction 1200 MeV p + $^{197}$Au. Experimental data by Budzanowski {\it et al.} \cite{Budzanowski} (green circles) are compared with results from CEM03.03F without coalescence expansion (blue solid lines) and CEM03.03F with coalescence expansion (red dashed lines). The coalescence expansion increases the production of high-energy $^6$He and $^6$Li, and improves agreement with experimental data. 

Fig.~\ref{fig:p480Ag} displays the fragment production spectra of $^7$Li and $^7$Be for the reaction 480 MeV p + $^{nat}$Ag. Experimental data by Green {\it et al.} \cite{Green480} (green circles) are compared with results from CEM03.03F without coalescence expansion (blue solid lines) and CEM03.03F with coalescence expansion (red dashed lines). Again, the coalescence expansion increases the production of heavy clusters, and improves agreement with experimental data. 

These reactions also highlight how the coalescence can produce heavy clusters not just of high-energy, but also of low- and moderate-energy, thus improving agreement with experimental data in these energy regions as well.

\begin{figure*}[t]
\centering
\includegraphics[width=6.0in]{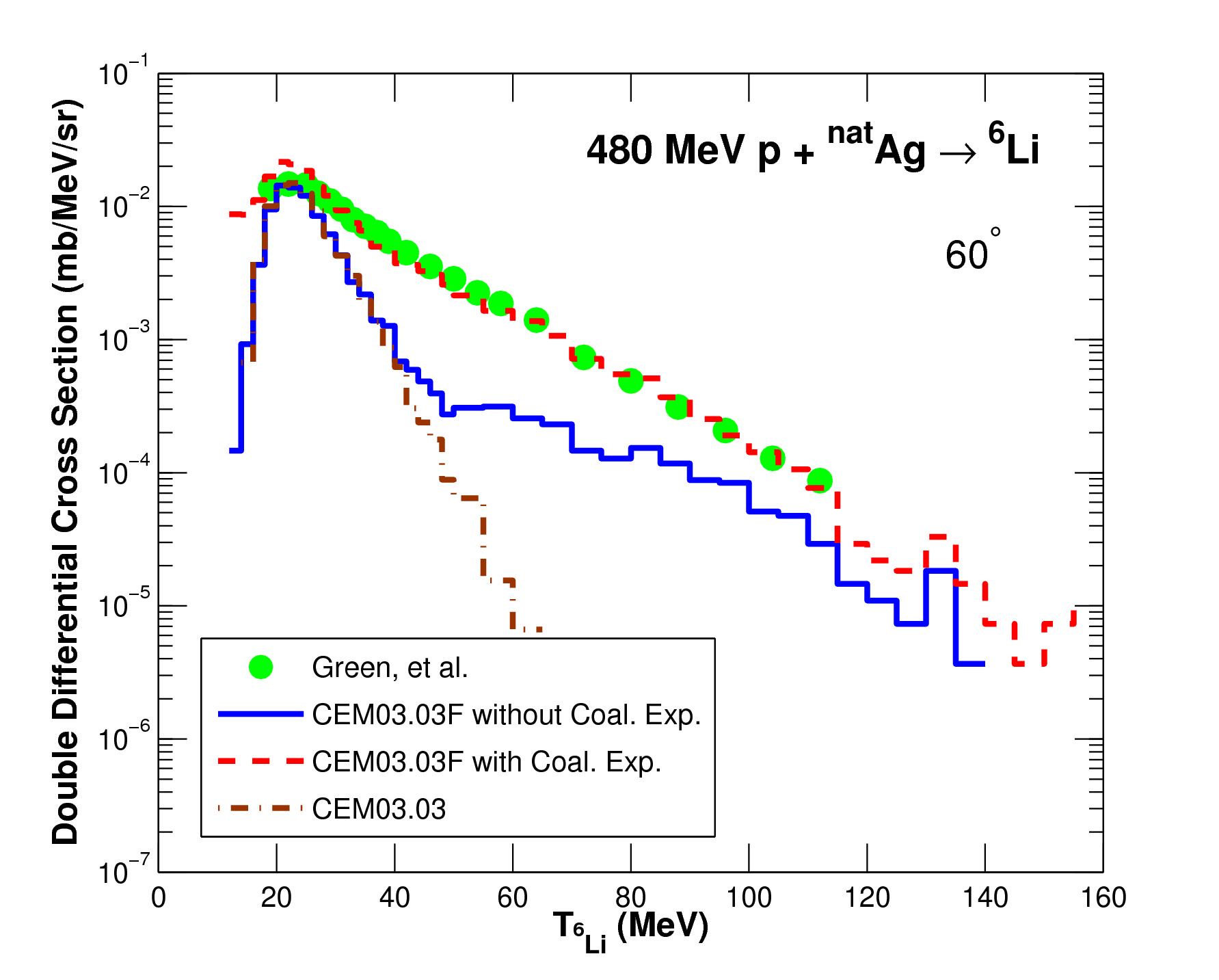}
\caption[]{Comparison of experimental results of the reaction 480 MeV p + $^{nat}$Ag $\rightarrow$ $^6$Li at 60\degree ~by Green {\it et al.} \cite{Green480} (green circles), with simulations from the original CEM03.03 (brown dashed-dotted lines), CEM03.03F without coalescence expansion (blue solid lines) and the CEM03.03F with coalescence expansion (red dashed lines).}
\label{fig:p480AgLi6}
\end{figure*}

Fig.~\ref{fig:p480AgLi6} displays experimental results of the reaction 480 MeV p + $^{nat}$Ag $\rightarrow$ $^6$Li by Green {\it et al.} \cite{Green480} (green circles), compared with simulations from results from CEM03.03F without coalescence expansion (blue solid lines), CEM03.03F with coalescence expansion (red dashed lines), and the original CEM03.03 (brown dashed-dotted lines). Even without the coalescence expansion, CEM03.03F (which contains a preequilibrium expansion and a total reaction cross section improvement) yields much better results than CEM03.03 without any of these improvements. Adding the coalescence expansion produces even better results.

%Additional 
Similar results
for many other reactions induced by protons, neutrons, and heavy ions
(the last are
simulated with LAQGSM03.03 \cite{Trieste08}, but with an extended
coalescence model as described in Ref. \cite{NUFRA2015abs})
and further discussions 
%of results 
can be find in Refs.~\cite{CoalescenceLANL, NUFRA2015abs}.

%%%%%%%%%%%%%%%%%%%%%%%%%%%%%%%%%%%%%%%%%%%%%%%%%%%%%%%%%%%%%%%%%%%%%%%%%%%%%%%%
\section{Conclusions}
Expanding the coalescence model within CEM yields increased production of heavy clusters in nuclear spallation reactions, particularly in the high-energy region, but also in the low- and moderate-energy regions. Preliminary results indicate this coalescence expansion yields improved agreement with experimental data. We plan to implement these upgrades into MCNP6.

%%%%%%%%%%%%%%%%%%%%%%%%%%%%%%%%%%%%%%%%%%%%%%%%%%%%%%%%%%%%%%%%%%%%%%%%%%%%%%%%
\section{Acknowledgments}
One of us (LMK) is grateful to
\begin{enumerate}
\item{Dr. Stepan Mashnik, for his continued mentoring and ample technical and scientific support and encouragement;}
\item{Drs. Avneet Sood, Larry Cox, Forrest Brown, and Tim Goorley and Los Alamos National Laboratory for the opportunity to study with some of the world's greatest experts in nuclear physics, particularly high-energy physics.}
\item{Dr. Akira Tokuhiro, for his continued support and expertise in serving as my thesis advisor.}
\end{enumerate}

This study was carried out under the auspices of the National Nuclear Security Administration of the U.S. Department of Energy at Los Alamos National Laboratory under Contract No. DE-AC52-06NA25396.

This work is supported in part (for L.M.K.) by the M. Hildred Blewett Fellowship of the American Physical Society, www.aps.org.

%%%%%%%%%%%%%%%%%%%%%%%%%%%%%%%%%%%%%%%%%%%%%%%%%%%%%%%%%%%%%%%%%%%%%%%%%%%%%%%%

%\bibliographystyle{ans}
%\bibliography{bibliography}
\end{document}